\documentclass[10pt]{article}

\usepackage{amstext,amsgen,latexsym,amsmath}
\usepackage{amstext,amssymb,amsfonts,latexsym}
\usepackage{theorem}
\usepackage{pifont}
\usepackage{graphics,epsfig}

 \newcommand{\bs}{\bigskip}
 \newcommand{\ms}{\medskip}
 \newcommand{\n}{\noindent}
 \newcommand{\s}{\smallskip}
 \newcommand{\hs}[1]{\hspace*{ #1 mm}}
 \newcommand{\vs}[1]{\vspace*{ #1 mm}}

 \newcommand{\setempty}{\mathrm{\O}}
 
 \newcommand{\nat}{\mathbb{N}}
 
 \newcommand{\integers}{\mathbb{Z}}
 \newcommand{\rational}{\mathbb{Q}}
 \newcommand{\complex}{\mathbb{C}}

 \newcommand{\field}{\mathbb{F}}

 \newcommand{\prob}{{\mathrm{Prob}}}

 \newcommand{\ie}{\textrm{i.e.},\hspace*{2mm}}
 \newcommand{\eg}{\textrm{e.g.},\hspace*{2mm}}

 \newcommand{\AAA}{{\cal A}}
 \newcommand{\CC}{{\cal C}}
 \newcommand{\FF}{{\cal F}}

 \newcommand{\GG}{{\cal G}}
 
 \newcommand{\UU}{{\cal U}}

 \newcommand{\p}{\mathrm{P}}
 
 \newcommand{\bpp}{\mathrm{BPP}}
 
 \newcommand{\espace}{\mathrm{ESPACE}}
 \newcommand{\eespace}{\mathrm{EESPACE}}
 \newcommand{\ppoly}{\mathrm{P\mbox{/}poly}}

 \newcommand{\eqp}{\mathrm{EQP}}
 \newcommand{\nqp}{\mathrm{NQP}}
 \newcommand{\bqp}{\mathrm{BQP}}
 \newcommand{\qma}{\mathrm{{QMA}}}

 \newcommand{\tally}{\mathrm{TALLY}}
 
 \newcommand{\poly}{\mathrm{poly}}
 \newcommand{\polylog}{\mathrm{polylog}}

  \newcommand{\qpoly}{\mathrm{Qpoly}}
 \newcommand{\qpolylog}{\mathrm{Qpolylog}}
 \newcommand{\qlog}{\mathrm{Qlog}}

 \newcommand{\cvector}[2]{\left( \begin{array}{c} #1 \\%
      #2 \end{array}\right)}

 \def\bbox{\vrule height6pt width6pt depth1pt}

\theoremstyle{plain}
\theoremheaderfont{\bfseries}
 \newtheorem{theorem}{Theorem}[section]
 \newtheorem{lemma}[theorem]{Lemma}
 \newtheorem{proposition}[theorem]{Proposition}
 \newtheorem{corollary}[theorem]{Corollary}

 {\theorembodyfont{\rmfamily}
 \newtheorem{definition}[theorem]{Definition}}
 {\theorembodyfont{\rmfamily} }
 {\theorembodyfont{\rmfamily} }

 \newenvironment{proof}{\par \noindent
            {\bf Proof. \hs{2}}}{\hfill$\Box$ \vspace*{3mm}}

 \newenvironment{proofof}[1]{\vspace*{5mm} \par \noindent
         {\bf Proof of #1.\hs{2}}}{\hfill$\Box$ \vspace*{3mm}}

 \newcommand{\ceilings}[1]{\lceil #1 \rceil}
 
 \newcommand{\pair}[1]{\langle #1 \rangle}

 \newcommand{\qubit}[1]{| #1 \rangle}

 \newcommand{\full}{\mathrm{Full}\mbox{-}}
 
 \newcommand{\ignore}[1]{}

\newif\ifnotesw\noteswtrue
   {\ifnotesw\marginpar[\hfill\(\top\)]{\(\top\)}\fi}%
      {\ifnotesw\marginpar[\hfill\(\bot\)]{\(\bot\)}\fi}
      
\newcommand{\mnote}[1]%
   {\ifnotesw\marginpar%
	  [{\scriptsize\begin{minipage}[t]{\marginparwidth}
	  \raggedleft#1%
		  \end{minipage}}]%
	  {\scriptsize\begin{minipage}[t]{\marginparwidth}
	  \raggedright#1%
		  \end{minipage}}%
    \fi}

\begin{document} 

\pagestyle{plain} 
\vs{6}
\begin{center}
{\Large {\bf Polynomial Time Quantum Computation with Advice}}
\footnote{This work was in part supported by the Natural Sciences and Engineering Research Council of Canada.} \bs\\
\begin{tabular}{c@{\hspace{20mm}}c} {\sc Harumichi Nishimura} & {\sc Tomoyuki Yamakami} \end{tabular}\ms\\ 
{School of Information Technology and Engineering} \\ {University of Ottawa, Ottawa, Ontario,
Canada K1N 6N5} 
\end{center} 

\paragraph{Abstract.} Advice is supplementary information that enhances the computational
power of an underlying computation. This paper focuses on advice that
is given in the form of a pure quantum state and examines the influence of such
advice on the behaviors of an underlying polynomial-time quantum
computation with bounded-error probability. 

\s
\n{\bf Key Words:} computational complexity, quantum circuit, advice function

\section{Prologue}\label{Sec:1}
Quantum computation has emerged to shape a future computational
paradigm based on quantum physics. To carry out a given task faster
and more precisely, it is also practical to supplement such a quantum
computation with a small piece of information beside an original
input. Ideally, such information should be succinct 
and given equally to all inputs of fixed size. The notion of such supplemental information,
under the name of ``advice,'' was first sought in a classical setting
by Karp and Lipton \cite{KL82} in the early 1980s.  Originally, Karp
and Lipton introduced the notion of advice to characterize non-uniform
models of computations, following the early work of Savage
\cite{Sav72} and Adleman \cite{Adl78} on non-uniform Boolean circuits.

In this paper, we consider polynomial-time bounded-error quantum
computations that take advice, which is given in the form of a pure
quantum state (referred to as {\em quantum advice}).  Of particular
interest are the languages recognized by polynomial-time quantum
computations with quantum advice under the condition that the quantum
computations should not err with probability more than $1/3$, provided
that the given advice is correct. The major difference from the
original definition of Karp and Lipton is that we do not impose any
condition on the acceptance probability of an underlying quantum
computation whenever advice is supplied incorrectly, since such
languages, when advice is limited to classical states (specially
called {\em classical advice}), establish a direct correspondence to
non-uniform quantum circuits as well as tally languages. For simplicity,
we use the notation $\bqp/^*\FF$ for the collection of aforementioned
languages with classical advice whose size is particularly 
described by functions in $\FF$ (in contrast with 
the Karp-Lipton advice class $\bqp/\FF$)
and we write $\bqp/^*\mathrm{Q}\FF$ for the quantum advice case, 
where prefix ``Q'' represents ``quantum.''

A central question on an advised computation is how to hide meaningful
information into advice and how to recover this information from the
advice with high accuracy. The key issue in this paper is an efficient
use of quantum advice, from which the strengths and limitations of
advised quantum computations follow. Using quantum fingerprinting
\cite{BCWW01}, we demonstrate that subpolynomial-size quantum advice
is more useful than classical advice of the same size. In contrast,
quantum information theory sometimes draws a clear limitation on how efficiently
we can hide information into quantum advice. Using quantum random
access coding (QRAC) \cite{ANTV02}, we show that quantum advice cannot be
made shorter than the 8 per cent of the size of classical advice 
for specific languages. Moreover, by combining the QRAC with our
quantum-circuit characterization, we construct a set in $\eespace$
that does not belong to $\bqp/^*\qpoly$. This result can be compared with 
Kannan's earlier result $\espace\nsubseteq\p/\poly$ \cite{Kan82}.

The use of quantum amplitudes is another way to enhance computational
power.  We can hide information within amplitudes and use a quantum
computation to access such information. Adleman, DeMarrais, and Huang
\cite{ADH97} showed that quantum computation can
benefit more from complex amplitudes than from rational amplitudes by 
proving $\bqp_{\rational}\neq\bqp_{\complex}$. This clearly contrasts
the recent result $\nqp_{\rational}=\nqp_{\complex}$ \cite{YY99}. To
some extent, we can view such complex amplitudes as advice given to an
underlying quantum computation with rational amplitudes. We show
that a finite set of complex amplitudes are roughly equivalent to
polylogarithmic advice.

We assume the reader's familiarity with the fundamental concepts in
the theory of computational complexity (\eg \cite{DK00}) and
quantum computation (\eg \cite{NC00}).  In this paper, all {\em
logarithms} have base 2 and a {\em polynomial} means a multi-variate 
polynomial with integer coefficients. 
We fix our alphabet $\Sigma$ to be $\{0,1\}$ unless otherwise stated. A {\em pairing function}
$\pair{\cdot,\cdot}$ is a map from $\Sigma^*\times\Sigma^*$ to
$\Sigma^*$, assumed to be one-to-one and polynomial-time computable
with the polynomial-time computable inverse. We also use the same
notation $\pair{\cdot,\cdot}$ for a standard bijection from
$\nat\times\nat$ to $\nat$, where $\nat=\{0,1,2,\cdots \}$.  
A {\em quantum string} ({\em qustring}, for short) {\em of length $n$} 
is a pure quantum state of $n$ qubits.
For any qustring $\qubit{\phi}$, $\ell(\qubit{\phi})$ denotes the length of
$\qubit{\phi}$. Let $\Phi_n$ be the collection of all qustrings of
length $n$ and define $\Phi_{\leq m}=\bigcup_{1\leq i\leq
m}\Phi_{i}$. The union $\bigcup_{n\in\nat}\Phi_n$ is denoted
$\Phi_\infty$ \cite{Yam99}. For convenience, let $\nat^+ =\nat-\{0\}$.

\section{Advice for Quantum Computation}\label{sec:advice} 

We focus on a polynomial-time quantum computation with bounded-error
probability as an underlying computation that takes advice.  We model
a quantum computation by a {\em multi-tape quantum Turing machine}
({\em QTM}, for short) whose heads are allowed to stay still
\cite{BV97,Deu85,ON00,Yam99}. Hereafter, the term ``QTM'' refers to a
QTM $M$ whose time-evolution is precisely described by a certain unitary
operator over the space spanned by all configurations of $M$.  For
convenience, our QTMs as well as classical TMs are 
equipped with multiple input tapes.
Whenever we write $M(x,y)$, we assume that $x$ is given 
in $M$'s {\em first}
input tape and $y$ is in the {\em second} input tape. 
Similarly, we use the notation $M(x,y,z)$ for three input tapes. 
Let $\prob_M[M(\qubit{\phi_1},\qubit{\phi_2})=s]$ 
denote the probability that a binary string $s$
is observed on the designated output tape of a QTM $M$ after 
$M$ halts on qustring inputs $\qubit{\phi_1}$ and $\qubit{\phi_2}$. 
When $M$'s amplitudes are concerned, we say that $M$
has {\em $K$-amplitudes} if all amplitudes of $M$ are chosen from a
subset $K$ of $\complex$.

Now, we want to define our central notion, a quantum advice
complexity class. To cope with the quantum nature of underlying
computations, we give the following special definition to our advice
class. The justification of our definition will be given in Section
\ref{sec:circuit}. For simplicity, we identify a set $A$ with its
characteristic function (\ie $A(x)=1$ if $x\in A$ and $0$ otherwise).

\begin{definition}\label{def:Qpoly} 
Let $f$ be any function from $\nat$ to $\nat$ and let $\FF$ be any set
of functions mapping from $\nat$ to $\nat$. Let $K$ be any nonempty
subset of $\complex$.

1. A set $A$ is in $\bqp_{K}/^*f$ (or $\bqp_{K}/^*f(n)$) if there
exist a polynomial-time QTM $M$ with $K$-amplitudes and a function $h$
from $\nat$ to $\Sigma^*$ such that $\prob_{M}[M(x,h(|x|))=A(x)]\geq
2/3$ for every $x\in\Sigma^*$, where $|h(n)|=f(n)$.  This function $h$
is called a {\em classical advice function} and $f$ is the {\em length
function} of $h$. Let $\bqp_{K}/^*\FF=\bigcup_{f\in\FF}\bqp_{K}/^*f$.

2. A set $A$ is in $\bqp_{K}/^*\mathrm{Q}f$ (or $\bqp_{K}/^*\mathrm{Q}(f(n))$) if there
exist a polynomial-time QTM $M$ with $K$-amplitudes and a function $h$
from $\nat$ to $\Phi_{\infty}$ such that $\ell(h(|x|))=f(|x|)$ and
$\prob_{M}[M(x,h(|x|))=A(x)]\geq 2/3$ for every $x\in\Sigma^*$, where
$h$ is called a {\em quantum advice function}.  Let $\bqp_{K}/^*\mathrm{Q}\FF
=\bigcup_{f\in\FF} \bqp_{K}/^*\mathrm{Q}f$.
\end{definition}

The prefix ``BQP$_{K}$'' in $\bqp_{K}/^*\FF$ and $\bqp_{K}/^*\mathrm{Q}\FF$ is an 
abbreviation of ``bounded-error quantum polynomial-time with $K$-amplitudes.'' Similar notions can be
introduced to probabilistic computations ($\bpp/^*\FF$) and other
types of quantum computations ($\eqp_{K}/^*\FF$ and
$\qma_{K}/^*\FF$). For readability, we suppress the subscript ``$K$''
if $K$ is the set of all {\em polynomial-time approximable complex
numbers} (that is, their real and imaginary parts are both
deterministically approximated to within $2^{-k}$ in time polynomial
in $k$). We are particularly interested in polynomial-length
and logarithmic-length functions. Conventionally, write
$\poly$ for the collection of all functions $f$ from $\nat$ to $\nat$
satisfying that $f(n)\leq p(n)$ for all $n\in\nat$, where $p$ is a
certain polynomial. Similarly, write $\log$ for the collection of all $f$'s
satisfying that $f(n)\leq c\log{n}+c$ for a certain fixed nonnegative
integer $c$.

Earlier, Karp and Lipton \cite{KL82} defined a general advice complexity
class\footnote{The Karp-Lipton advice class $\CC/\FF$ is the
collection of all sets $A$ for which there exist a set $B\in\CC$,
a function $f\in\FF$, and a function $h$ from $\nat$ to $\Sigma^*$
such that $A=\{x\mid \langle x,h(|x|) \rangle\in B\}$ provided that
$|h(n)|=f(n)$ for all $n\in\nat$.} $\CC/\FF$ for any class $\CC$
of languages and any set $\FF$ of length functions.  This Karp-Lipton
style definition naturally introduces another advice class $\bqp/\FF$ for
the {\em language class} $\bqp$ of Bernstein and Vazirani
\cite{BV97}.  Clearly, $\bqp/\FF$ is included in $\bqp/^*\FF$ for any
set $\FF$ of length functions.  The major difference between
$\bqp/^*\FF$ and $\bqp/\FF$ is that the definition of $\bqp/^*\FF$
lacks the promise-free property of underlying QTMs, where a QTM $M$ is called
{\em promise-free} if, for every pair $(x,s)$, either
$\prob_{M}[M(x,s)=0]\geq 2/3$ or $\prob_{M}[M(x,s)=1]\geq 2/3$. Such a
difference seems, nonetheless, insignificant in the classical setting
since the corresponding two definitions $\bpp/^*\poly$ and $\bpp/\poly$
coincide\footnote{Let $A\in\bpp/^*\poly$. By a standard majority vote technique, 
there exist a polynomial $p$, a polynomial-time deterministic TM $M$, and a polynomial advice function $h$ such that $\prob_{r\in\Sigma^{p(|x|)}}[M(x,h(|x|),r)=A(x)]\geq 1-2^{-2n}$ for all $x$. 
For each $n$, choose an $r_n\in\Sigma^{p(n)}$ that satisfies $M(x,h(n),r_n)=A(x)$ for all $x\in\Sigma^n$. 
By setting the new advice $k(n)=0^{p(n)}1h(n)r_n$, we obtain $A\in\ppoly\subseteq\bpp/\poly$.}.  
We note that there is no known proof for the collapse between
$\bqp/^*\poly$ and $\bqp/\poly$. Their separation on the contrary seems
difficult to prove since $\mathrm{Promise}\mbox{-}\p=\mathrm{Promise}\mbox{-}\bqp$ implies
$\p/\FF=\bqp/\FF = \bqp/^*\FF$, where $\mathrm{Promise}\mbox{-}\CC$ is
the promise version of complexity class $\CC$ \cite{EY80}.

The following fundamental properties hold for advice classes $\bqp/^*\FF$ and $\bqp/^*\mathrm{Q}\FF$.  
The {\em power set} of $\Sigma^*$ is denoted $2^{\Sigma^*}$ in the lemma below.

\begin{lemma}\label{basic}
Let $f$ and $g$ be any functions from $\nat$ to $\nat$ and let $\FF$
and $\GG$ be any sets of functions from $\nat$ to $\nat$. 

(1) $\bqp/^*0 = \bqp/^*\mathrm{Q}(0) = \bqp$. 

(2) $\bqp/^*2^n = \bqp/^*\mathrm{Q}(2^n) = 2^{\Sigma^*}$.

(3) $\bqp/^*\FF \subseteq \bqp/^*\mathrm{Q}\FF$.

(4) If $\FF\subseteq \GG$ then $\bqp/^*\FF\subseteq \bqp/^*\GG$ and
$\bqp/^*\mathrm{Q}\FF \subseteq \bqp/^*\mathrm{Q}\GG$.

(5) If $g(n)<f(n)\leq 2^n$ for infinitely many $n$, then
$\p/^*f\nsubseteq \bqp/^*g$.
\end{lemma}

The complexity class $\bqp$ is known to enjoy a strong form of the
so-called {\em amplification property}, for which we can amplify the
success probability of any underlying QTM from $2/3$ to $1-2^{-p(n)}$
for an arbitrary polynomial $p$. This form of the amplification
property can be easily extended into any classical advice class
$\bqp/^*\FF$. The quantum advice
class $\bqp/^*\mathrm{Q}\FF$, 
however, demands a more delicate attention since
quantum advice in general cannot be copied due to the {\em no-cloning
theorem}. For the following lemma, we say that a set $\FF$ of length
functions is {\em closed under integer multiplication} if, for
every $f\in\FF$ and every integer $k\in\integers$, there exists a function
$g\in\FF$ such that $f(n)\cdot k\leq g(n)$ for all $n\in\nat$.

\begin{lemma}\label{amplification}{\rm (Amplification Lemma)} 
Let $\FF$ be any set of length functions.  
(1) A set $A$ is in $\bqp/^*\FF$ if and only if, for every polynomial
$q$, there exist a polynomial-time QTM $M$ and a classical advice
function $h$ whose length function is in $\FF$ such that
$\prob_{M}[M(x,h(|x|))=A(x)]\geq 1-2^{-q(|x|)}$ for every $x$. 
(2) Assume that $\FF$ is closed under integer multiplication. A set
$A$ is in $\bqp/^*\mathrm{Q}\FF$ if and only if, for every constant
$\epsilon\geq0$, there exist a polynomial-time QTM $M$ and a quantum advice
function $h$ whose length function is in $\FF$ such that
$\prob_{M}[M(x,h(|x|))=A(x)]\geq 1-\epsilon$ for every $x$.
\end{lemma}

\section{Non-Uniform Quantum Circuits and Tally Sets}\label{sec:circuit}

Our definition $\bqp/^*\FF$ is preferable to the Karp-Lipton style
definition $\bqp/\FF$ because, as shown in Lemma \ref{equiv}, our
definition can precisely characterize non-uniform polynomial-size
quantum circuits, where a {\em quantum circuit} \cite{Deu89,Yao93} is
assumed to be built from a finite universal set of quantum gates and
the {\em size} of a quantum circuit is the number of quantum gates in
use.

Throughout this paper, we fix a {\em universal} set $\UU$ of quantum
gates, consisting of a Controlled-NOT gate and a finite number of
single-qubit gates that generates a dense subset in $SU(2)$ 
with their inverses. Without loss of generality, 
we may assume that all entries of these quantum gates
are polynomial-time approximable complex numbers. We say that a set $A$
has {\em non-uniform polynomial-size quantum circuits with error
probability $\epsilon$} if there exist a polynomial $p$ and a
non-uniform family $\{C_n\}_{n\in\nat}$ of quantum circuits such that,
for every string $x$, (i) $C_{|x|}$ on input $\qubit{x}\qubit{0^m}$
outputs $A(x)$ with probability at least $1-\epsilon$, where $\qubit{0^m}$ is
an auxiliary input and (ii) $C_{|x|}$ uses at most $p(|x|)$ quantum
gates chosen from $\UU$. The notation $\prob_{C}[C(x,y)=b]$ expresses
the probability that $C$, taking $x$ and $y$ as a pair of inputs with
an auxiliary input $0^m$, outputs $b$ to the first qubit of
$C$.

\begin{lemma}\label{equiv} 
(1) A set $A$ is in $\bqp/^*\poly$ if and only if $A$ has non-uniform
polynomial-size quantum circuits with error probability at most $1/3$. 

(2) A set $A$ in $\bqp/^*\qpoly$ if and only if there exist a positive polynomial
$p$, a non-uniform family $\{C_n\}_{n\in\nat}$ of polynomial-size
quantum circuits, and a series $\{U_n\}_{n\in\nat}$ of unitary
operators acting on $p(n)$ qubits such that, for every $n\in\nat$ 
and every string
$x$ of length $n$, $\prob_{C_n}[C_n(x,U_n\qubit{0^{p(n)}})=A(x)]\geq 2/3$.
\end{lemma}

Obviously, Lemma \ref{equiv}(1) is a special case of (2).  The ``only if''
part of Lemma \ref{equiv}(2) follows from the explicit simulation of
QTMs by quantum circuits \cite{NO02,Yao93}. For any set $A$ in
$\bqp/^*\qpoly$ via a polynomial quantum advice function $h$, we can
build a family $\{C_n\}_{n\in\nat}$ of polynomial-size quantum
circuits such that $\prob_{C_n}[C_n(x,h(n))=A(x)]\geq 2/3$ for every
$x$ of length $n$. The ``if'' part needs an effective binary
encoding of a quantum circuit, provided that the length of such an
encoding is not less than the size of the circuit.  We use the
notation $Code(C)$ to describe this encoding of a quantum circuit
$C$. If a quantum circuit $C_n$ satisfies
$\prob_{C_n}[C_n(x,U_n\qubit{0^{p(n)}})=A(x)]\geq 2/3$ for every $x$
of length $n$, the desired advice function $h(n)$ is defined to be the
encoding $Code(C_n)$ tensored with the qustring $U_n\qubit{0^{p(n)}}$. 
This puts $A$ into $\bqp/^*\qpoly$ via $h$. The above characterizations in Lemma \ref{equiv} 
represent the clear difference between $\bqp/^*\qpoly$ and $\bqp/^*\poly$ because $U_n\qubit{0^{p(n)}}$ 
may be exponentially difficult to construct. However, it does not address the separation 
between $\bqp/^*\poly$ and $\bqp/^*\qpoly$.

The following lemma allows us to replace the unitary operator $U_n$ in
Lemma \ref{equiv}(2) by any exponential-size quantum circuit with no
ancillary qubit. This lemma can be obtained directly from the
Solovay-Kitaev theorem \cite{Kit97,NC00} following the standard
decomposition of unitary matrices \cite{Bar95}. 
For any complex square matrix $A$, let
$\|A\|=\sup_{\qubit{\phi}\neq0}\|A\qubit{\phi}\|/\|\qubit{\phi}\|$.

\begin{lemma}\label{qustring-gate} 
(1) For every sufficiently large $k\in\nat$, every
$\qubit{\phi}\in\Phi_k$, and every $\epsilon>0$, there exists a
quantum circuit $C$ acting on $k$ qubits such that $C$ has size at
most $2^{2k}\log^3{(1/\epsilon)}$ and
$\|C\qubit{0^k}-\qubit{\phi}\|<\epsilon$. 

(2) For every sufficiently large $k\in\nat$, every $k$-qubit
unitary operator $U_k$, and every $\epsilon>0$, there exists a quantum
circuit $C$ acting on $k$ qubits such that $C$ has size at most
$2^{3k}\log^3{(1/\epsilon)}$ and $\|U(C)-U_k\|<\epsilon$, where $U(C)$
is the unitary operator representing $C$.
\end{lemma}

The quantum-circuit characterization of $\bqp/^*\poly$ yields the
following containment.

\begin{proposition}\label{bqpqlogvsbqppoly}
$\bqp/^*\qlog \subseteq \bqp/^*\poly$.
\end{proposition}

\begin{proof} Assume that $A\in\bqp/^*\qlog$. 
By Lemma \ref{amplification}(2), there exist a polynomial-time QTM $M$ and a series
$\{\qubit{\psi_n}\}_{n\in\nat}$ of qustrings of length logarithmic in
$n$ such that $\prob_{M}[M(\qubit{x},\qubit{\psi_{|x|}}) \neq
A(x)]\leq 1/6$ for every string $x$.  There exists a family
$\{C_n\}_{n\in\nat}$ of polynomial-size quantum circuits that
simulate $M$. By Lemma \ref{qustring-gate}(1), each
$\qubit{\psi_n}$ can be approximated to within $1/6$ by a certain
quantum circuit $D_n$ of size polynomial in $n$. Combining $C_n$ with
$D_n$ produces a new quantum circuit of polynomial size that
recognizes $A\cap\Sigma^{n}$. This implies that $A$ has
polynomial-size quantum circuits with error probability at most
$1/6+1/6 = 1/3$. By Lemma \ref{equiv}(1), $A$ is
in $\bqp/^*\poly$.
\end{proof}

Non-uniform quantum circuits also characterize polylogarithmic advice
classes. For each positive integer $k$, let $\log^k$ be the collection
of all functions $f$ from $\nat$ to $\nat$ such that $f(n)\leq
c(\log{n})^k+c$ for any $n\in\nat$, where $c$ is a certain fixed nonnegative
integer.  In early 1990s, Balc{\'a}zar, Hermo, and Mayordomo
\cite{BHM92} showed that $\p/\log^k$ can be expressed in terms of
Boolean circuits whose encodings belong to the resource-bounded
Kolmogorov complexity class $K[\log^k,\poly]$, which is the collection
of all languages $A$ such that any string $x$ in $A$ can be produced
deterministically in time polynomial in $|x|$ from a certain string
$w$ (called a {\em program}) of length at most $f(|x|)$ for a certain
function $f\in \log^k$. The following lemma naturally expands their
result into $\bqp/^*\log^k$.

\begin{lemma}\label{kolmogorov}
Let $k\in\nat^{+}$. A set $A$ is in $\bqp/^*\log^k$ if and only if
there is a non-uniform family $\{C_n\}_{n\in\nat}$ of polynomial-size
quantum circuits that recognize $A$ with probability $\geq 2/3$ and
satisfy $\{Code(C_n)\mid n\in\nat\}\in K[\log^k,\poly]$.
\end{lemma} 

Notice that a polynomial-size quantum circuit can be encoded into 
a set of strings of polynomial length over a single-letter alphabet. 
Hence, there is a strong connection between polynomial-size quantum circuits
and tally sets, where a {\em tally} set is a subset of $\{0\}^*$ or
$\{1\}^*$. In particular, the collection of all tally sets is
represented as $\tally$. Using Lemma \ref{equiv}(1), we can establish
the following tally characterization of $\bqp/^*\poly$, which expands
the classical result $\p/\poly=\p^{\tally}$ \cite{BH77}. This lemma
also supports the legitimacy of our definition $\bqp/^*\FF$.

\begin{lemma}\label{tt1} 
$\bqp/^*\poly = \bqp^{\tally}$.
\end{lemma}

The tally characterization of a logarithmic advice class draws special
attention. Unlike $\bqp/^*\poly$, $\bqp/^*\log$ is not closed even under
polynomial-time Turing reductions (P-T-reductions, for short) since
$\p^{\bqp/^*\log}=\bqp/^*\poly$ but $\bqp/^*\poly\neq\bqp/^*\log$. Because
of a similar problem on $\p/\log$, Ko \cite{Ko87} gave an alternative
definition to a logarithmic advice class, which is now known as $\full\p/\log$
\cite{BH98,BHM92}. (Ko \cite{Ko87} originally called this class 
$\mathrm{Strong}\mbox{-}\p/\log$.) Similarly, we introduce the new advice class
$\full\bqp/^*\log$.

\begin{definition}\label{def:full}
Let $f$ be any length function. A set $A$ is in $\full\bqp/^*f$ if
there exist a polynomial-time QTM $M$ and a function $h$ from $\nat$
to $\Sigma^*$ such that, for all $n$, $|h(n)|=f(n)$ and
$\prob_{M}[M(x,h(n))=A(x)]\geq 2/3$ for any string $x$ of length at
most $n$. For a class $\FF$ of length functions, let $\full\bqp/^*\FF$
denote the union $\bigcup_{f\in\FF}\full\bqp/^*f$.
\end{definition}

It is clear from Definition \ref{def:full} that
$\full\bqp/^*\FF\subseteq \bqp/^*\FF$ for any set $\FF$ of length
functions. Note that $\full\bqp/^*\poly=\bqp/^*\poly$.
Let $\mathrm{TALLY2}$ denote
the collection of all subsets of $\{0^{2^k} \mid k\in\nat\}$
\cite{BH98}.

\begin{lemma}\label{tt2}
$\full\bqp/^*\log = \bqp^{\tally2}$.
\end{lemma}

The proof of Lemma \ref{tt2} is a straightforward modification of the
proof for $\full\p/\log=\p^{\tally2}$ \cite{BH98,BHM92}. It immediately
follows from the lemma that $\full\bqp/^*\log$ is closed under
P-T-reductions. 

\section{Power of Quantum Advice}\label{sec:strength}

To make efficient use of quantum advice, we want to embed classical
information schematically into shorter quantum advice and retrieve the
information using quantum computation with small errors. The following
theorem implies that subpolynomial quantum advice is more useful than
classical advice of the same size. For the theorem, we introduce the
following terminology: a function $f$ from $\nat$ to $\nat$ is called
{\em infinitely-often polynomially bounded} if there is a positive 
polynomial
$p$ such that $f(n)\leq p(n)$ for infinitely-many numbers $n$ in
$\nat$.

\begin{theorem}\label{fingerprinting}
Let $f$ be any positive length function. If $f$ is infinitely-often
polynomially bounded, then 
$\bqp_{K}/^*\mathrm{Q}(O(f(n)\log{n}))\nsubseteq
\bqp/^*f(n)\cdot n$, where $K=\{0,1\}$.
\end{theorem}

By choosing an appropriate $f$ in Theorem \ref{fingerprinting}, we
obtain the following consequence. The union of $\log^k$ for all
$k\in\nat^{+}$ is denoted $polylog$.

\begin{corollary} 
(1) $\bqp/^*\log\neq \bqp/^*\qlog$. 

(2) $\bqp/^*n^k\neq \bqp/^*\mathrm{Q}(n^k)$ 
    for each fixed $k\in\nat^{+}$.  

(3) $\bqp/^*\qlog\nsubseteq \bqp/^*\polylog$ and hence, 
$\bqp/^*\polylog\neq \bqp/^*\qpolylog$.
\end{corollary}

To prove Theorem \ref{fingerprinting}, we use the notion of quantum
fingerprinting introduced by Buhrman, Cleve, Watrous, and de Wolf
\cite{BCWW01}. The following simple quantum
fingerprint given in \cite{Wol01} suffices for our proof. Fix
$n$ and $\epsilon>0$. Let $\field_{n,\epsilon}$ be any field of size
$pw(n/\epsilon)$, where $pw(m)$ is the least prime power larger than
$m$.  Note that $pw(n/\epsilon)\leq 2n/\epsilon$.  For any string
$x=x_1\cdots x_n$ of length $n$, the {\em quantum fingerprint} $\qubit{\phi_n(x)}$ of $x$ is the
qustring of length $2\ceilings{\log(pw(n/\epsilon))}$ defined by
$\qubit{\phi_n(x)}=(1/\sqrt{|\field_{n,\epsilon}|})\sum_{z\in\field_{n,\epsilon}}
\qubit{z}\qubit{p_x(z)}$, where $p_x(z)$ denotes the polynomial
$p_x(z)= \sum_{i=1}^{n}x_i\cdot z^{i-1}$ over $\field_{n,\epsilon}$.  

\begin{proofof}{Theorem \ref{fingerprinting}}
Fix an arbitrary positive polynomial $p$ satisfying $f(n)\leq p(n)$ for
infinitely-many $n$ in $\nat$. Assume an effective enumeration of
polynomial-time QTMs, say $\{M_i\}_{i\in\nat^{+}}$. 
We construct by stages the
set $L$ that separates $\bqp_{K}/^*\mathrm{Q}(O(f(n)\log{n}))$ from
$\bqp/^*f(n)n$. At stage $0$, let $n_0=0$.  At stage $i\geq1$, choose
the minimal integer $n_i$ such that $n_i>n_{i-1}$, $f(n_i)\leq
p(n_i)$, and $n_i> 2(1+\log{p(n_i)})$. Consider the collection
$C_{n_i}$ of all sets $A\subseteq\Sigma^{n_i}$ that satisfy the
following criterion: there exists a string $s\in\Sigma^{f(n_i)n_i}$
satisfying $\prob_{M_i}[M_i(x,s)=A(x)]\ge 2/3$ for all
$x\in\Sigma^{n_i}$. Note that there are at most $2^{f(n_i)n_i}$ such
sets. By contrast, there are exactly $\sum_{j=0}^{2f(n_i)}{\tiny
\cvector{2^{n_i}}{j}}$ subsets of $\Sigma^{n_i}$ of cardinality at
most $2f(n_i)$. Since $\sum_{j=0}^{2f(n_i)}{\tiny
\cvector{2^{n_i}}{j}}> (2^{n_i}/2f(n_i))^{2f(n_i)}
> 2^{f(n_i)n_i}\geq |C_{n_i}|$, we can find a set $L_{n_i} \subseteq
\Sigma^{n_i}$ of cardinality $\leq 2f(n_i)$ that does not 
belong to $C_{n_i}$.  Take such a set $L_{n_i}$ for each
$i\in\nat^{+}$ and define $L=\bigcup_{i\ge 1}L_{n_i}$.  Since
$L_{n_i}\not\in C_{n_i}$ for all $i\in\nat^{+}$, $L$ is located
outside $\bqp/^*f(n)n$. 

To complete the proof, we show that $L$ is in $\bqp/^*\mathrm{Q}(f(n)\log{n})$. 
Write $k(n)$ for $2f(n)n$. Fix $i$ and write $n$ for $n_i$ for readability. Take a field $\field_{k(n),1/4}$ and
define $g(n)$ $=$ 
$\qubit{0^m1}\qubit{\phi_{k(n)}(y_{1})}\qubit{\phi_{k(n)}(y_{2})}
\cdots\qubit{\phi_{k(n)}(y_{m})}$ when $L_n=\{y_1,y_2,\ldots,y_m\}$ 
for a certain number $m\leq 2f(n)$. Recall that
$|\field_{k(n),1/4}|\geq 8nf(n)$. Consider the following algorithm
$\AAA$: given input $(x,g(|x|))$, if, for some $i\in\{1,2,\ldots,m\}$,
the first half part of $\qubit{\phi_{k(n)}(y_i)}$ is $z$ and $p_x(z)$
equals the second half part of $\qubit{\phi_{k(n)}(y_i)}$, then accept
the input. If there is no such $i$, then reject the input.

Now, take any string $x$ of length $n$.  Clearly, if $x\in L_n$, then
$\AAA$ always accepts the input in time polynomial in $n+g(n)$.  If
$x\not\in L_n$, then $p_x\neq p_{y_{j}}$ for any
$j\in\{1,\ldots,m\}$. Since $p_x$ and $p_{y_{j}}$ have degree at most
$n-1$, they agree on at most $n-1$ elements in $\field_{k(n),1/4}$.
Thus, the probability that $\AAA$ erroneously accepts the input is at
most $m\cdot(n-1)/|\field_{k(n),1/4}| <1/4$. Overall, we can recognize $L$ with error probability at most $1/4$ in polynomial
time. Since $f(n)\leq p(n)$, the length of quantum advice $g(n)$ is at
most $f(n)+1+ 2\ceilings{f(n)\log(pw(4f(n)n))} \leq cf(n)\log{n}+c$,
where $c$ is an appropriate constant independent of $n$.  
Hence, we have $L\in \bqp_{K}/^*\mathrm{Q}(O(f(n)\log{n}))$.
\end{proofof}

A careful examination of the above proof suggests that 
a deterministic Turing machine with ``randomly selected'' classical advice of size $f(n)n$ 
can replace a QTM with quantum advice of the same size. 
The difference between randomly selected advice and quantum advice is open 
while the former can be simulated by the latter. 

\section{Limitation of Quantum Advice}\label{sec:limit}

Quantum fingerprinting demonstrates in Section \ref{sec:strength} an
efficient way to compress a large volume of classical information into
relatively-short quantum advice. There is, however, a quantum
information theoretical limitation on such quantum compression. In the
following theorem, we claim that quantum advice cannot be made shorter
than classical advice with the multiplicative factor of at least 0.08 on the QTMs. 

\begin{theorem}\label{qrac}
For any positive length function $f$ such that $f(n)\leq 2^n$,  
$\p/f\nsubseteq \bqp/^*\mathrm{Q}(0.08f(n))$. 
\end{theorem}

Theorem \ref{qrac} contrasts with the result
$\p/f\nsubseteq\bqp/^*(f(n)-1)$ obtained from Lemma \ref{basic}(5). As
a consequence of Theorem \ref{qrac}, we can show the following
corollary.

\begin{corollary}  
(1) $\p/\log^2\nsubseteq \bqp/^*\qlog$. 

(2) $\p/\poly\nsubseteq\bqp/^*\mathrm{Q}(n^k)$ 
   for each fixed $k\in\nat^{+}$ and 
   hence $\bqp/^*\qlog\neq \bqp/^*\poly$. 

(3) $\bqp/^*\qlog\subsetneq \bqp/^*\mathrm{Q}(n^k) 
   \subsetneq\bqp/^*\qpoly$ 
   for each fixed $k\in\nat^{+}$.
\end{corollary}

The proof of Theorem \ref{qrac} requires a lower bound of quantum
random access encodings introduced by Ambainis, Nayak, Ta-shma, and Vazirani \cite{ANTV02}. 
An {\em $(n,m,p)$-quantum random access coding} (QRAC) is a function $f$ that maps $n$-bit strings to (pure or
mixed) quantum states over $m$ qubits satisfying the following: for 
every $i\in\{1,\ldots,n\}$, there is a measurement $O_i$ with outcome
0 or 1 such that $\prob[O_i(f(x))=x_i]\ge p$ for all $x\in\Sigma^n$. 
It is known in \cite{ANTV02} that any $(n,m,p)$-QRAC should
satisfy the inequality $m\ge (1-H(p))n$, where $H(p)=-p\log p-(1-p)\log(1-p)$. 
We now prove Theorem \ref{qrac}. 

\begin{proofof}{Theorem \ref{qrac}}
Let $\{M_i\}_{i\in\nat^{+}}$ be any effective
enumeration of polynomial-time QTMs. We
build by stages the set $L=\bigcup_{n\in\nat}L_n$ that separates $\p/\FF$ from $\bqp/^*\mathrm{Q}(0.08f(n))$.  
At stage $n$,
consider the set $\AAA_n$ of all subsets of $\{x\in\Sigma^n\mid \text{$x$ is lexicographically at most $s_{f(n)}$}\}$, 
where $s_i$ is lexicographically the $i$th string in
$\Sigma^n$. Note that $\AAA_n$ can be viewed as the set of all
strings of length $f(n)$. For each $s\in\Sigma^{f(n)}$, let $B_s$ be 
the set in $\AAA_n$ such that $s=B_s(s_1)B_s(s_2)\cdots B_s(s_{f(n)})$. 
Consider any number $m\geq1$ satisfying the following: for every $s\in\Sigma^{f(n)}$,
there exists a qustring $\qubit{\phi_s}\in\Phi_{m}$ such that
$\prob_{M_n}[M_n(x,\qubit{\phi_s})=B_s(x)]\geq 2/3$ for all
$x\in\Sigma^n$. Since the function $g$ defined as $g(s)=\qubit{\phi_s}$ for
every $s\in\Sigma^{f(n)}$ is an $(f(n),m,2/3)$-QRAC, we obtain
$m\geq (1-H(1/3))f(n) > 0.08f(n)$ for
all $n$. Therefore, there exists a string
$s\in\Sigma^{f(n)}$ such that no qustring $\qubit{\phi}$ in
$\Phi_{m}$, where $m\leq 0.08f(n)$, satisfies
$\prob_{M_n}[M_n(x,\qubit{\phi})=B_s(x)]\geq2/3$ for all
$x\in\Sigma^n$. Choose such an $s$ and define $L_n=B_s$. The above
construction guarantees that $L\not\in \bqp/^*\mathrm{Q}(0.08f(n))$. 
Since $|L_n|\leq f(n)$ for all $n$, we have $L\in \p/f$.
\end{proofof}

Another application of QRACs yields the
existence of a set in $\eespace$ that does not belong to
$\bqp/^*\qpoly$, where $\eespace$ is the class of all sets computed by
deterministic Turing machines using $2^{2^{O(n)}}$ space. Similarly,
$\espace$ is defined using $2^{O(n)}$ space.

\begin{theorem}\label{espace} 
(1) $\espace\nsubseteq \bqp/^*\poly$. 

(2) $\eespace\nsubseteq \bqp/^*\qpoly$. 
\end{theorem}

Theorem \ref{espace}(1) expands Kannan's result 
$\espace\nsubseteq \ppoly$ \cite{Kan82}. The proof of
Theorem \ref{espace}(2) combines a diagonalization argument with the lower bound of QRACs.

\begin{proofof}{Theorem \ref{espace}} 
We show only 2) since 1) can be obtained by an argument similar 
to the proof of $\espace\nsubseteq \ppoly$ \cite{Kan82}. 
Let $\{M_i\}_{i\in\nat^{+}}$ be any effective enumeration of all
polynomial-time QTMs and let $\{p_i\}_{i\in\nat^{+}}$ be that of all
polynomials with nonnegative coefficients. 
Note from Lemma \ref{qustring-gate}(1) that any qustring
of length $m$ can be approximated to within $1/6$ by a certain quantum
circuit with input $\qubit{0^m}$ of size at most $2^{2m+6}$. Consider
the following algorithm $\AAA$ that starts with the empty input and
proceeds by stages.
\begin{quote}
{\sf At stage $0$, set $Q=\setempty$. At stage $n\geq1$, first
enumerate all numbers in $\{1,2,\ldots,n\}\setminus Q$ in the
increasing order. For each of such numbers $m$, we carry out the
following procedure. At round $m=\pair{i,j}$, for each quantum circuit
$D$ of size at most $2^{2p_i(n)+6}$ acting on $p_i(n)$ qubits, 
compute $z_D^{(m)}=z_1\cdots z_{2^n}$ as follows. For each $k$ ($1\leq k\leq
2^n$), let $z_k$ be the outcome (either $0$ or $1$) of $M_j$ on input
$(s_k,D\qubit{0^{p_i(n)}})$ with probability $\geq 5/6$, where
$s_k$ is lexicographically the $k$th string in $\Sigma^n$. If some
$z_k$ does not exist, then let $z_D^{(m)}$ be undefined and go to next
$D$. After all $D$'s are examined, consider the set $Z$ of all
$z_D^{(m)}$'s (which are defined). If both $Z=\Sigma^{2^n}$ and $m<n$,
then go to next round $m+1$. Assume otherwise. If $m=n$ then output
$\bot$, or else output the minimal $z$ not in $Z$ and let
$Q=Q\cup\{m\}$. Go to next stage $n+1$. }
\end{quote}
Now, we show that $Q$ eventually equals $\nat$. Assume otherwise. Let
$m=\pair{i,j}$ be the minimal number not in $Q$. Take any sufficiently
large number $n_0$ and assume that, at any stage $n\geq n_0$, $\AAA$
always checks $M_j$ at its first round. This happens when $\Sigma^{2^n}$ 
equals the set of all $z_D^{(m)}$'s for all $n\geq n_0$. Hence, for
every length $n\geq n_0$ and every set $A\subseteq\Sigma^n$, there
exists a qustring $\qubit{\phi_{n,A}}$ of length $p_i(n)$ such that
$\prob_{M_i}[M_i(x,\qubit{\phi_{n,A}})=A(x)]\geq 2/3$ for all
$x\in\Sigma^n$. Letting $A[n]=A(0^n)A(0^{n-1}1)\cdots A(1^n)$ for each
$n$, we define $f(A[n])=\qubit{\phi_{n,A}}$. Since $f$ is a
$(2^n,p_i(n),2/3)$-QRAC, it follows that 
$p_i(n)\geq (1-H(1/3))2^n >0.08\cdot 2^n$ for all $n\geq n_0$, 
a contradiction. Therefore, $Q=\nat$.

The desired language $L$ is defined as follows: $x\in L$ if and only
if $\AAA$ outputs a binary string whose $k$th bit is 1 at stage $|x|$,
assuming that $x$ is the $k$th string in $\Sigma^{|x|}$. Then, $L$ is not in $\bqp/^*\qpoly$ 
since, otherwise, $L$ is recognized by a certain QTM $M_j$ with quantum advice of length $p_i(n)$ 
with high probability, and thus $\pair{i,j}\not\in Q$, which contradicts $Q=\nat$. 
On the other hand, $L\in\eespace$ since $L(x)$ is computed by running $\AAA$ up to stage $|x|$ using space $2^{O(2^{|x|})}$.
\end{proofof}
\section{Roles of Amplitudes as Advice}\label{sec:amplitude}

Amplitudes can be viewed as a resource given to quantum
computations. We can hide meaningful information within amplitudes and
recover it using a certain type of quantum computation. Adleman,
DeMarrais, and Huang \cite{ADH97} first demonstrated how to hide such
information and proved that $\bqp_{\complex}$ properly includes
$\bqp$, which equals $\bqp_{\rational}$.  We further claim that
amplitudes may play a role of logarithmic advice. 

\begin{theorem}\label{full-complex}
$\full\bqp/^*\log \subseteq \bqp_{\complex} \subseteq \bqp/^*\log^3$. 
\end{theorem}

\begin{proof}
The first inclusion is shown in the following fashion. It is
sufficient to prove that $\mathrm{TALLY2}\subseteq\bqp_{\complex}$
since $\full\bqp/^*\log = \bqp^{\tally2}\subseteq
\bqp^{\bqp_{\complex}}=\bqp_{\complex}$ by Lemma \ref{tt2}.  Assume
that $L$ is any set in $\tally2$. We encode $L$ into the real number
$\theta_L =2\pi(\sum_{n=1}^{\infty}
\frac{h(n)}{8^n})$, where $h(n)=(-1)^{1-L(0^{2^n})}$. 
Consider the QTM $M$ that carries out the following algorithm: 
given input $x$, reject $x$ if $x\neq 0^{2^k}$ for any
$k\in\nat$.  If $x=0^{2^k}$ for $k=\log{|x|}$, 
then prepare the state $\qubit{0}$,
conduct the transformation $\qubit{0}\mapsto
\cos(8^{k-1}\theta_L+\pi/4)\qubit{0}
+\sin(8^{k-1}\theta_L+\pi/4)\qubit{1}$, and 
measure it on the $\{\qubit{0},\qubit{1}\}$-basis.
If the result of the measurement is 1, then 
accept $x$, or else reject $x$. 
An argument similar to the proof of Theorem 5.1 in \cite{ADH97}
shows that, on any input $x$, $M$ outputs $L(x)$ in polynomial time
with probability at least $2/3$. This concludes that $L$ is indeed in
$\bqp_{\complex}$. 
The second inclusion is shown as follows. Let $L$ be any set in
$\bqp_{\complex}$ recognized by a polynomial-time QTM $M$ 
with error probability $\leq 2^{-n}$ together with its amplitudes chosen from
$\complex$. Let $p$ be any polynomial that bounds the running time of
$M$. Since the transition function of $M$ is a finite function, it
induces the corresponding unitary operator acting over a
finite-dimensional Hilbert space.  Let $U(M)$ denote this unitary
operator. By choosing $k=\mathrm{dim}(U(M))$ and $\epsilon=1/n^{c}$ 
for a sufficiently large $c>0$ in Lemma \ref{qustring-gate}(2), 
we obtain a family of quantum circuits
$\{C_n\mid n\in\nat\}$ of size $O(\log^3{n})$ such that each $C_n$
implements a unitary matrix $U(C_n)$ satisfying $||U(C_n)-U(M)||\le
1/3p(n)$.  Note that all single-qubit gates in $C_n$ have
polynomial-time approximable numbers as their components. With the help
of the encoding $Code(C_n)$ as an advice string, we can simulate $M$
with error probability at most $p(n)\cdot (1/3p(n)) \leq 1/3$ in
polynomial time.  This implies that $L$ is in $\bqp/^*\log^3$.
\end{proof}

Theorem \ref{full-complex} leads to the following direct consequence. 

\begin{corollary}\label{amplitude}
(1) $\bqp_{\complex}\subsetneq \bqp/^*\polylog$. 

(2) $\bqp_{\rational}/^*\FF=\bqp_{\complex}/^*\FF$ and $\bqp_{\rational}/^*\mathrm{Q}\FF
=\bqp_{\complex}/^*\mathrm{Q}\FF$ for any $\FF\in\{\polylog,\poly\}$.
\end{corollary}

The proof of Corollary \ref{amplitude}(2) needs the fact that
$\bqp_{\rational}/^*\FF=\bqp/^*\FF$ and
$\bqp_{\rational}/^*\mathrm{Q}\FF=\bqp/^*\mathrm{Q}\FF$ for any 
set $\FF$. In Theorem \ref{full-complex}, 
however, we cannot replace $\full\bqp/^*\log$ by $\bqp/^*\log$ or even $\bqp/^*1$.

\begin{proposition}
$\bqp/^*1 \nsubseteq \bqp_{\complex}$ and thus, $\bqp/^*\log\nsubseteq
\bqp_{\complex}$.
\end{proposition}

\begin{proof}
Assume that $\bqp/^*1\subseteq\bqp_{\complex}$. Recall from Lemma
\ref{tt1} that $\bqp/^*\poly=\bqp^{\tally}$. Since
$\tally\subseteq\bqp/^*1$, it follows that
$\bqp/^*\poly=\bqp^{\tally}\subseteq
\bqp^{\bqp/^*1}\subseteq\bqp^{\bqp_{\complex}}=\bqp_{\complex}$. 
Hence, we obtain $\bqp_{\complex}=\bqp/^*\poly$, which contradicts 
Corollary \ref{amplitude}(1).   
\end{proof}

\section{Epilogue}

We have initiated a study of advised quantum computations and addressed several relations 
among complexity classes with classical or quantum advice and some known complexity classes. 
The important questions we left open include: 
(1) $\bqp/^*\poly=?\,\bqp/^*\qpoly$, (2) $\bqp\subseteq?\,\eqp/^*\qpoly$, 
and (3) $\espace\nsubseteq?\,\bqp/^*\qpoly$. 



\end{document}